\documentclass[aps,prb,amsmath,amssymb,floatfix,twocolumn,amsmath,superscriptaddress,showkeys,showpacs]{revtex4-1}
\usepackage{color,epsfig}
\usepackage{bm}
\usepackage{amsmath,amsfonts,amssymb,bm}
\usepackage{hyperref}
\usepackage{subfigure}
\usepackage{feynmf}

\newcommand{\bea}{\begin{eqnarray}}
\newcommand{\eea}{\end{eqnarray}}
\newcommand{\be}{\begin{equation}}
\newcommand{\ee}{\end{equation}}

\newcommand{\sgn}{{\rm sign}}

\renewcommand\Re{\text{Re}}

\begin{document}

\title{Chemical sensing with graphene: A quantum field theory perspective}
\author{Horacio Falomir}
\affiliation{Facultad de Ciencias Exactas de la UNLP, IFLP, CONICET—Departamento de Física, C.C. 67, (1900) La Plata, Argentina}
\author{Marcelo Loewe}
\affiliation{Facultad de F\'isica, Pontificia Universidad Cat\'olica de Chile, Vicu\~{n}a Mackenna 4860, Santiago, Chile}
\affiliation{Centro Cient\'ifico Tecnol\'ogico de Valpara\'iso, CCTVAL, Universidad T\'ecnica Federico Santa Mar\'ia, Casilla 110-V, Valpara\'iso, Chile}
\affiliation{Centre for Theoretical and Mathematical Physics, University of Cape Town, Rondebosch 770, South Africa}
\author{Enrique Mu{\~n}oz}
\email{munozt@fis.puc.cl}
\affiliation{Facultad de F\'isica, Pontificia Universidad Cat\'olica de Chile, Vicu\~{n}a Mackenna 4860, Santiago, Chile}
\affiliation{Center for Nanotechnology and Advanced Materials CIEN-UC, Avenida Vicuña Mackenna 4860, Santiago, Chile}

\date{\today}

\begin{abstract}
We studied theoretically the effect of a low concentration of adsorbed polar molecules on the optical conductivity of graphene, within the Kubo linear response approximation. Our analysis is based on
a continuum model approximation that includes up to next to nearest neighbors in the pristine graphene effective Hamiltonian,
thus extending the field-theoretical
analysis developed in Refs.\cite{Falomir_2018,Falomir_2020}.  Our results show that the
conductivity can be expressed in terms of renormalized
quasiparticle parameters $\tilde{v}_F$, $\tilde{M}$ and $\tilde{\mu}$ that include the effect of the molecular
surface concentration $n_{dip}$ and dipolar moment $\boldsymbol{\mathcal{P}}$, thus providing an analytical model for a graphene-based chemical sensor. 
\end{abstract}

\maketitle

\section*{Introduction}
The remarkable transport properties of graphene\cite{DasSarma_11,CastroNeto_09,Peres_10,Munoz_012,Munoz_010}, as well as its affinity for the physisorption of different molecules, has attracted much attention towards its application as a field-effect transistor (FET) in chemical sensing\cite{An_2020,Azizi_2020,Gao_2018,Yavari_2012}. In particular, the detection of
polar molecules in gas phase has been investigated both experimentally as well as theoretically\cite{Zhang_2015,Zhang_2009,Chen_2010,Yavari_2012}, with the later approach 
mainly based on ab-initio methods. While providing an accurate prediction of the electronic structure for single adsorbed molecules\cite{Appelt_2018}, ab-initio methods are not
suitable to describe molecular concentrations, finite temperature and disorder effects.
On the other hand, in complement with those numerical studies, the effects of molecular concentration, disorder and finite temperature can be described by analytical models based on the continuum Dirac approximation within quantum field theory\cite{Vozmediano_10,deJuan_12,Arias_015,Falomir_2018}, thus providing an intuitive and accurate\cite{Peeters2011} picture of the
underlying physical phenomena. Moreover, with appropriate approximations, these analytical models can often provide explicit formulae that are useful to interpret actual experiments. In this work, 
we present an analytical theory for the optical conductivity in graphene under a given concentration of adsorbed polar molecules. This theory is a direct application of our previous work \cite{Falomir_2018,Falomir_2020}, based on
a continuum description of graphene involving the
effects of up to next-to-nearest neighbors on the
underlying atomistic tight-binding Hamiltonian \cite{Falomir_2012}, that is analyzed by means of quantum field theory methods to include the electrostatic effects of adsorbed polar molecules on the surface of graphene. We assume that the spatial distribution, as well as the orientation of the dipole moments are disordered. The continuum model describing pristine graphene
is summarized by the Lagrangian density\cite{Falomir_2018,Falomir_2020}
\begin{eqnarray}
\mathcal{L} &=&
\frac{i}{2}\left[ {\psi}^{\dagger}\partial_{t}\psi - \partial_{t}\psi^{\dagger}\psi \right]
+\psi^{\dagger}e A_0\psi -\frac{1}{2M}
\left\{\left[\left(\mathbf{p}-e\mathbf{A}\right.\right.\right.\nonumber\\
&&\left.\left.\left.+\theta\boldsymbol{\sigma}\right)\psi\right]^{\dagger}\cdot
\left[\left(\mathbf{p}-e\mathbf{A} + \theta\boldsymbol{\sigma}\right)\psi\right]
-2\theta^2\psi^{\dagger}\psi
\right\},
\label{eq:Lagrangian}
\end{eqnarray}
with $\theta = M v_F$, and the effective mass
parameter\cite{Falomir_2018,Falomir_2020,Falomir_2012} $M = -2/(9 t' a_0^2)<0$ capturing the
effect of next-to-nearest neighbor hopping $t'$ on the continuum energy spectrum near both Dirac points.
\begin{figure}
\centering
    \includegraphics[width=0.46\textwidth]{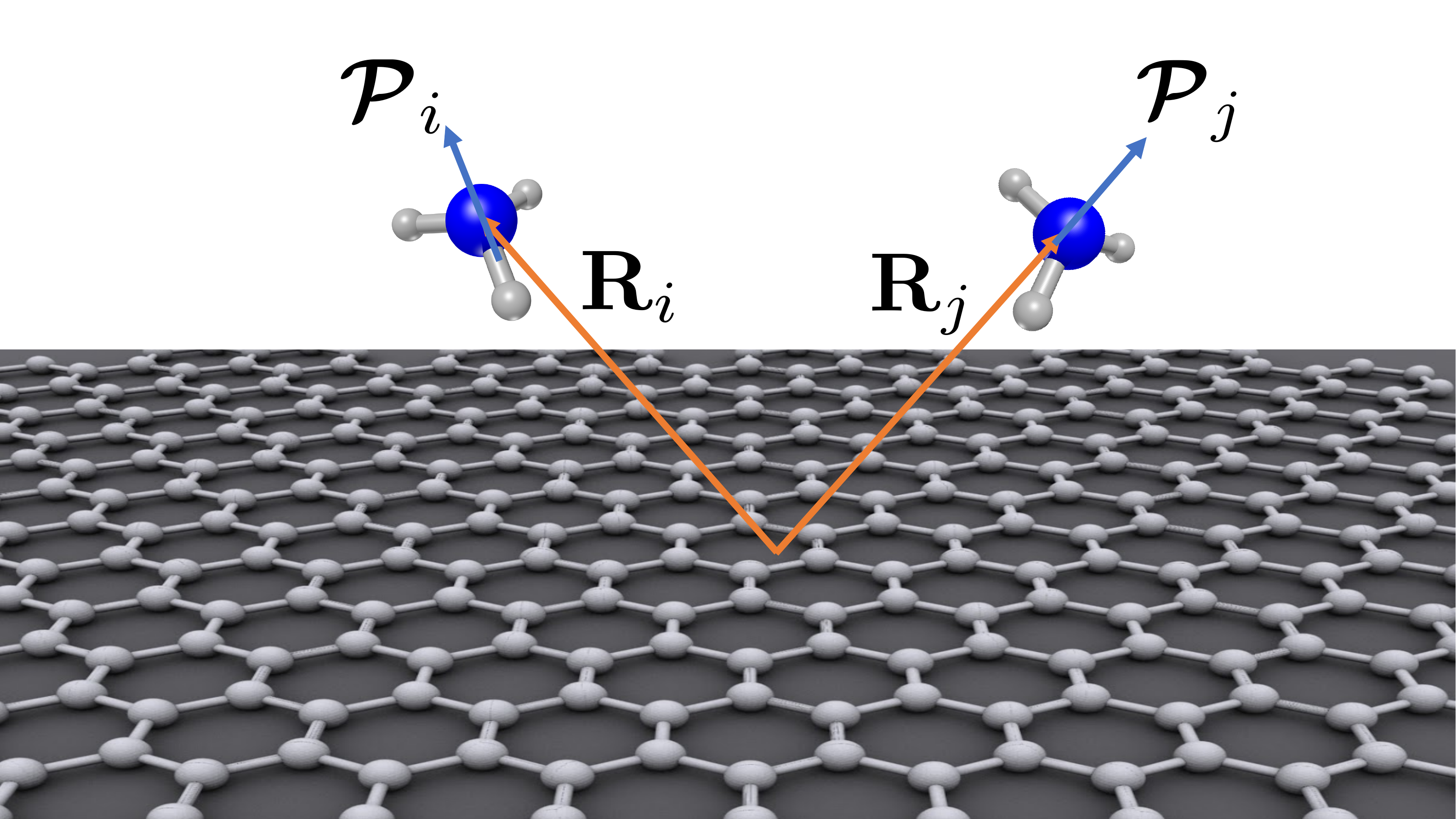}
    \caption{Pictorial (not in actual scale) representation of polar molecules adsorbed at
    positions $\mathbf{R}_i$ and $\mathbf{R}_j$ on the surface of graphene.}
    \label{fig:Molec}
\end{figure}

\section*{Scattering by randomly adsorbed molecules}
We assume for simplicity that the electric properties of a single molecule adsorbed at a distance $a>0$ above the surface of graphene, can be modeled through a dipole potential, with two point charges $+Q$ and $-Q$, located at $(\mathbf{d}/2,a)$ and $(-\mathbf{d}/2,a)$, respectively, with $\mathbf{d}$ a two-dimensional vector. The correspondig potential at a position $\mathbf{r} = (x,y,0)$ on the
surface of graphene ($z = 0$) is
\begin{eqnarray}
V_{\mathbf{d}}(\mathbf{r}) &= \frac{\mathcal{P}}{4\pi\epsilon d}\left(
\frac{1}{\sqrt{(\mathbf{r} - \mathbf{d}/2)^2 +  a^2}}- \frac{1}{\sqrt{(\mathbf{r} + \mathbf{d}/2)^2  + a^2}}
\right)
\label{eq:pot}
\end{eqnarray}
with $\boldsymbol{\mathcal{P}} = Q\mathbf{d}$ the dipole moment of the molecule and $\epsilon$ the local dielectric permittivity. The strict dipole approximation corresponds to the limit
$d\rightarrow 0$, while $\boldsymbol{\mathcal{P}}$ remains finite. In this limit, the 2D
Fourier transform of the potential in Eq.(\ref{eq:pot}) reduces to (see Appendix~\ref{AppendixA})
\begin{eqnarray}
\hat{V}_{\boldsymbol{\mathcal{P}}}(\mathbf{q}) &=& \lim_{d\rightarrow 0}\hat{V}_{\mathbf{d}}(\mathbf{q})
= \frac{i \boldsymbol{\mathcal{P}} \cdot\mathbf{q}}{2\epsilon q }
e^{- q a}
\label{eq:dip2}
\end{eqnarray}
\begin{figure}
    \includegraphics[width=0.46\textwidth]{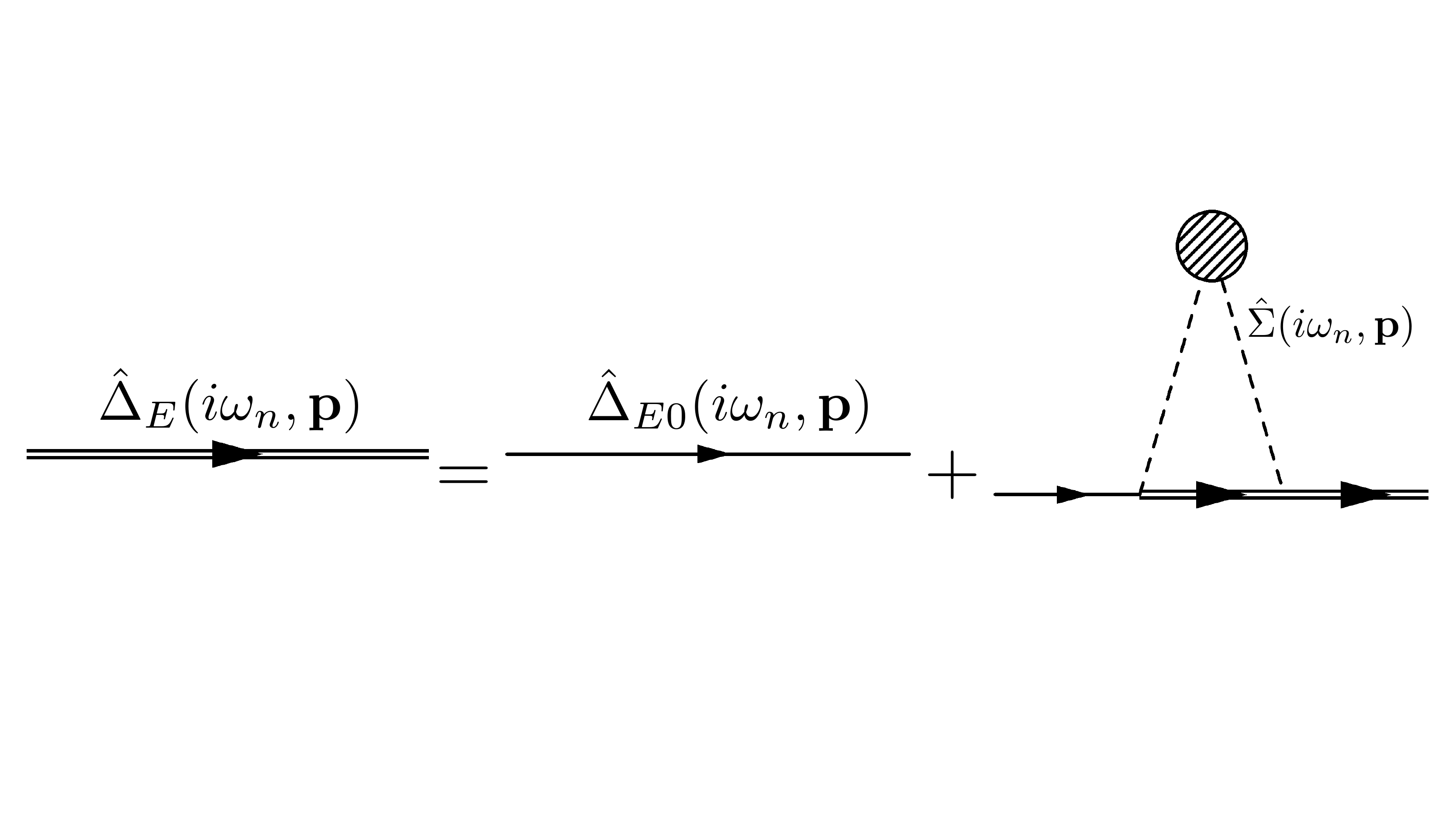}
    \caption{The Feynman diagram corresponding to the Dyson equation for the dressed propagator. The self-energy is expressed by Eq.~(\ref{eq:selfenergy})}.
    \label{fig:Feyn}
\end{figure}
Let us now consider the effect of scattering by adsorbed polar molecules, randomly distributed with a surface concentration $n_{dip}$, on the
optical conductivity of graphene. For this purpose, we first analyze their effect on the
effective propagators arising from the Lagrangian Eq.~(\ref{eq:Lagrangian}) at finite temperature. A given realization within the ensemble of adsorbed molecular configurations is given by the potential
\begin{eqnarray}
V(\mathbf{r}) = \sum_{j = 1}^{N}V_{\boldsymbol{d}_{j}}(\mathbf{r} - \mathbf{R}_j).
\end{eqnarray}
The basic assumptions are that the positions $\mathbf{R}_j$,
as well as the orientations of the dipolar moments $\boldsymbol{\mathcal{P}}_{j}$ are independent and identically distributed random variables, i.e.
\begin{eqnarray}
\langle \mathbf{R}_i\cdot\mathbf{R}_j \rangle = \delta_{ij} \langle \mathbf{R}^2\rangle,\,\,
\langle \boldsymbol{\mathcal{P}}_{i}\cdot\boldsymbol{\mathcal{P}}_{j} \rangle = \delta_{ij} \frac{\mathcal{P}^2}{2},
\end{eqnarray}
where in the last equation we further assumed that the dipole orientations are uniformly distributed in the azimuthal angle.
Using standard diagrammatic methods and average over disorder\cite{Rammer_86}, it is shown (see Appendix~\ref{AppendixB} for details) that for
low molecular concentrations $n_{dip}$, the scattering effects are correctly described by 
a disorder-averaged self-energy of the form (see Appendix~\ref{AppendixA})
\begin{eqnarray}
\hat{\Sigma}(i\omega_n,\mathbf{p}) = n_{dip}\int\frac{d^2 q}{\left(2\pi \right)^2}
|\hat{V}_{avg}(\mathbf{p}-\mathbf{q})|^2\hat{\Delta}_E(i\omega_n,\mathbf{q}).
\label{eq:selfenergy}
\end{eqnarray}
Here, $\hat{\Delta}_E(\omega_n,\mathbf{p})$ is the fully dressed propagator, arising from the solution of the Dyson equation (as depicted in Fig.\ref{fig:Feyn})
\begin{eqnarray}
\hat{\Delta}^{-1}_E(i\omega_n,\mathbf{p}) = \hat{\Delta}^{-1}_{E0}(i\omega_n,\mathbf{p}) - \hat{\Sigma}(i\omega_n,\mathbf{p}),
\label{eq:Dyson}
\end{eqnarray}
while the Fourier transform of the dipole potential, averaged
over dipole orientations is defined as (see Appendix~\ref{AppendixA})
\begin{eqnarray}
|\hat{V}_{avg}(\mathbf{p})|^2 = \frac{1}{2} \frac{e^2\mathcal{P}^2}{4 \epsilon^2}e^{-2 p a}.
\end{eqnarray}
From our previous work \cite{Falomir_2018,Falomir_2020}, the bare, finite temperature (Euclidean) propagator is given by the expression
\begin{eqnarray}
\hat{\Delta}_{E0}^{-1}(i\omega_n,\mathbf{p}) = \left(i\omega_n + \mu - \frac{\mathbf{p}^2}{2M} \right)\mathbf{1} - v_F \boldsymbol{\sigma}\cdot\mathbf{p},
\label{eq:Eprop0}
\end{eqnarray}
where the effective mass parameter involved in the quadratic dispersion term arises from the next to nearest neighbors hopping $t'$ in the atomistic tight-binding Hamiltonian of graphene \cite{Falomir_2012}, $M = -2/(9 a_0^2 t')\sim -1.36\cdot10^{-30}$~Kg. 
On the other hand, we show that the self-energy is given by the expression (see Appendix~\ref{AppendixB})
\begin{eqnarray}
\hat{\Sigma}(i\omega_n,\mathbf{p}) &=& n_{dip}\frac{e^2\mathcal{P}^2}{8\epsilon^2}
\left[\mathbf{1}\,\mathcal{I}_1(i\omega_n,\mathbf{p})+ \frac{\boldsymbol{\sigma}\cdot\mathbf{p}}{p}\mathcal{I}_2(i\omega_n,\mathbf{p}) \right]\nonumber\\
\label{eq:selfenergy2}
\end{eqnarray}
where the exact definitions of the scalar functions $\mathcal{I}_1(i\omega_n,\mathbf{p})$ and $\mathcal{I}_2(i\omega_n,\mathbf{p})$
are given in Appendix~\ref{AppendixB}. Consistently with the second nearest-neighbor contribution in the graphene Hamiltonian, we consider only contributions up to second order in momentum in these integrals, such that
\begin{eqnarray}
\mathcal{I}_1(i\omega_n,\mathbf{p}) &=& \mathcal{I}_1^{(0)}(i\omega_n) + \mathbf{p}^2\mathcal{I}_1^{(2)}(i\omega_n) + O(\mathbf{p}^4)\nonumber\\
\mathcal{I}_2(i\omega_n,\mathbf{p}) &=& p \mathcal{I}_2^{(1)}(i\omega_n) +  O(\mathbf{p}^3)
\end{eqnarray}
Inserting the expression for the self-energy Eq.(\ref{eq:selfenergy2}) and the bare propagator Eq.(\ref{eq:Eprop0}) into the Dyson Eq.(\ref{eq:Dyson}), we obtain the inverse full propagator
\begin{widetext}
\begin{eqnarray}
\hat{\Delta}^{-1}_E
&=& \left( i\omega_n + \mu - \frac{n_{dip}e^2\mathcal{P}^2}{8\epsilon^2} \mathcal{I}_1^{(0)}(i\omega_n) - \frac{\mathbf{p}^2}{2 M}\left[1 + \frac{ M n_{dip}e^2\mathcal{P}^2}{4\epsilon^2} \mathcal{I}_1^{(2)}(i\omega_n)\right]\right)\mathbf{1}-\boldsymbol{\sigma}\cdot\mathbf{p}\,v_F\left(
1 + \frac{n_{dip}e^2\mathcal{P}^2}{8\epsilon^2 v_F}\mathcal{I}_2^{(1)}(i\omega_n)
\right)\nonumber\\
&=& z^{-1}\left\{\left( i\omega_n + \tilde{\mu} - \frac{\mathbf{p}^2}{2 \tilde{M}}\right)\mathbf{1}
-\boldsymbol{\sigma}\cdot\mathbf{p}\,\tilde{v}_F
-\tilde{\Sigma}^{(2)}(\omega_n,\mathbf{p})
\right\}.
\label{eq:fullprop}
\end{eqnarray}
\end{widetext}

\begin{figure}
\centering
    \includegraphics[width=0.46\textwidth]{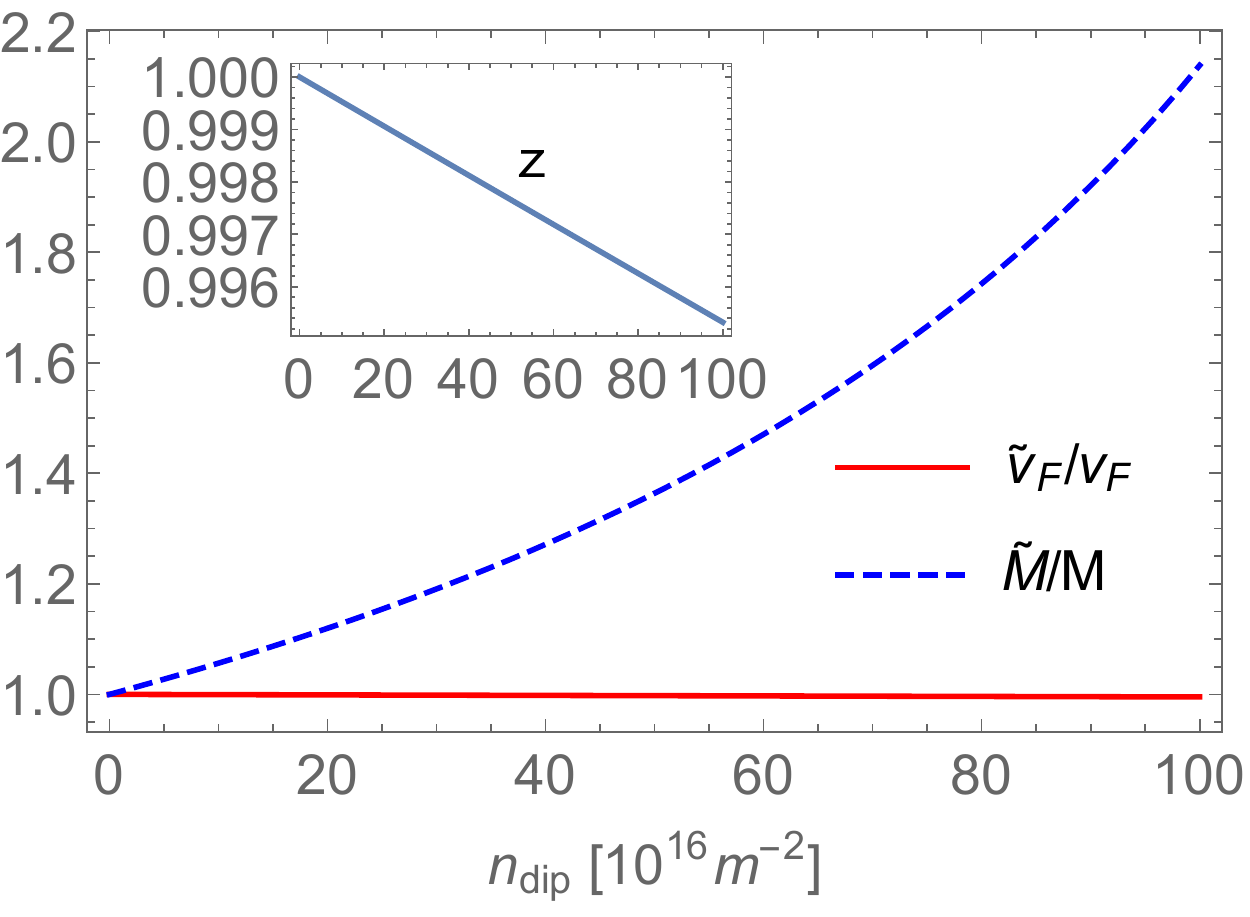}
    \caption{(Color online) Renormalized parameters $\tilde{M}/M$ (blue, dashed line) and $\tilde{v}_F/v_F$ (red, solid line), and renormalization factor $z$ (inset) as a function of the concentration of adsorbed polar molecules $n_{dip}$ (in units of $10^{16}m^{-2}$). }
    \label{fig:renparam}
\end{figure}

The fact that the tensor structure of the disorder-averaged propagator Eq.~(\ref{eq:fullprop}) is the same as the free one Eq.~(\ref{eq:Eprop0}), supports the renormalized quasiparticle picture \cite{Hewson_2001,Munoz_013}. Therefore, after expanding at low frequencies with respect to the chemical potential, we defined the renormalized quasiparticle parameters
\begin{eqnarray}
\tilde{M}^{-1} &=& z M^{-1}\left(1 +  \frac{ M n_{dip}e^2\mathcal{P}^2}{4\epsilon^2} \Re\,\mathcal{I}_1^{(2)}(0) \right)\nonumber\\
\tilde{\mu} &=& z\left(\mu - \frac{n_{dip}e^2\mathcal{P}^2}{8\epsilon^2} \Re\,\mathcal{I}_1^{(0)}(0) \right)\nonumber\\
\tilde{v}_F &=& z v_F \left(1 + \frac{n_{dip}e^2\mathcal{P}^2}{8\epsilon^2 v_F}\Re\,\mathcal{I}_2^{(1)}(0) \right),
\label{eq:renorparam}
\end{eqnarray}
with the wavefunction renormalization factor
\begin{equation}
z^{-1} = 1 - \frac{n_{dip}e^2\mathcal{P}^2}{8\epsilon^2}\left.\Re\,\frac{\partial}{\partial(i\omega)}\mathcal{I}_1^{(0)}\right|_{\omega=0}, 
\label{eq:z}
\end{equation}
while the matrix $\tilde{\Sigma}^{(2)}(\omega_n,\mathbf{p})$ contains the self-energy contributions at higher frequencies $O(\omega^2)$. Neglecting such higher energy contributions, we therefore define the  quasiparticle propagator
\begin{eqnarray}
\tilde{\Delta}_{E}^{-1}(i\omega_n,\mathbf{p})
=\left( i\omega_n + \tilde{\mu} - \frac{\mathbf{p}^2}{2 \tilde{M}}\right)\mathbf{1}
-\boldsymbol{\sigma}\cdot\mathbf{p}\,\tilde{v}_F.
\label{eq:quasiprop}
\end{eqnarray}
The dependence of the renormalized quasiparticle parameters defined in Eq.~(\ref{eq:renorparam}) on the adsorbed molecular concentration
is presented in Fig.~\ref{fig:renparam}, where for illustration we have
chosen the parameters for $\text{NH}_3$, with a dipole moment of $\mathcal{P} = 1.42$~Debye, and $a = 3.6$~\AA.

\begin{figure}[b!]
\centering
    \includegraphics[width=0.46\textwidth]{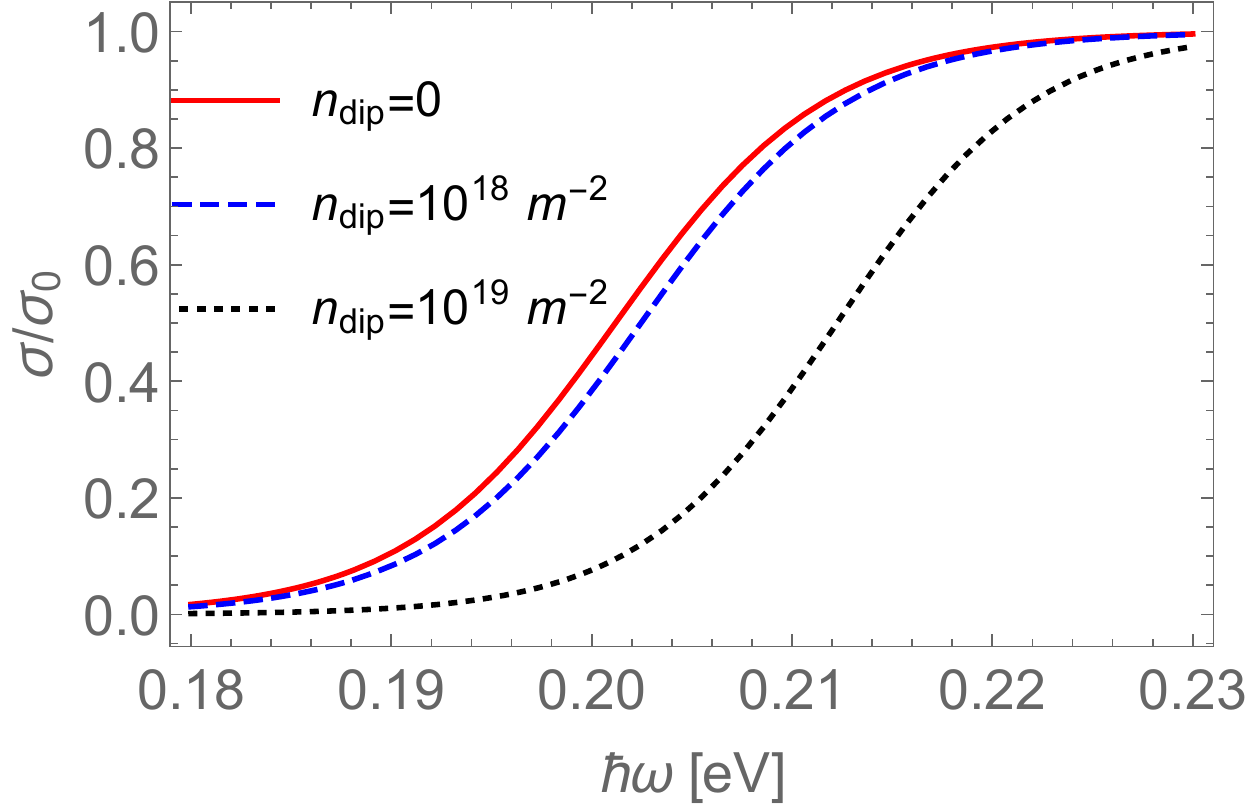}
    \caption{(Color online) Optical conductivity of graphene, calculated after Eq.~(\ref{eq:graphcond}) and normalized by the "universal" value $\sigma_0 = e^2/(4\hbar)$, as a function of
    frequency, for different adsorbed molecular concentrations $n_{dip}$. The temperature was set to $30$~K for this example.}
    \label{fig:cond}
\end{figure}

\section*{Conductivity}
From the Kubo linear response theory, the optical conductivity
tensor is given by the expression \cite{Falomir_2018,Falomir_2020}
\begin{eqnarray}
\sigma_{kl}(\omega) = 4 \times \left.\frac{\Pi_{kl}^{R}(p)}{i\omega}\right|_{p = (\omega,0)}
\label{eq:Kubo}
\end{eqnarray}
The retarded component of the polarization tensor is obtained via
analytic continuation from the Euclidean, following the standard prescription $\Pi_{kl}^{R}(\omega,\mathbf{p}) = \Pi_{kl}^{E}(i\omega_n\rightarrow \omega + i \epsilon,\mathbf{p})$.

The Euclidean polarization tensor, within the quasiparticle approximation, is given by \cite{Falomir_2018,Falomir_2020}
\begin{eqnarray}
\Pi_{kl}^E(i \omega_n,\mathbf{p})
&=& \frac{e^2}{4 \tilde{M}^2}\frac{1}{\beta}\sum_{q_4=\omega_n,n\in\mathbb{Z}}\int\frac{d^2 q}{(2\pi)^2}\Gamma_{ab}^k
(p + 2 q)\nonumber\\
&&\times\tilde{\Delta}_{bc}^{E}(p+q)\Gamma_{cd}^{l}(p + 2q)\tilde{\Delta}_{da}^{E}(q),
\label{eq:pol}
\end{eqnarray}
where the integral involves the quasiparticle propagators defined by Eq.(\ref{eq:quasiprop}), and we defined the matrices \cite{Falomir_2018}
\begin{eqnarray}
\Gamma_{ab}^{k}(p) = \left[ \delta_{ab} p^k + 2\tilde{M}\tilde{v}_F \left[ \sigma^k\right]_{ab}  \right].
\label{eq:gam}
\end{eqnarray}

\begin{figure}
\centering
    \includegraphics[width=0.46\textwidth]{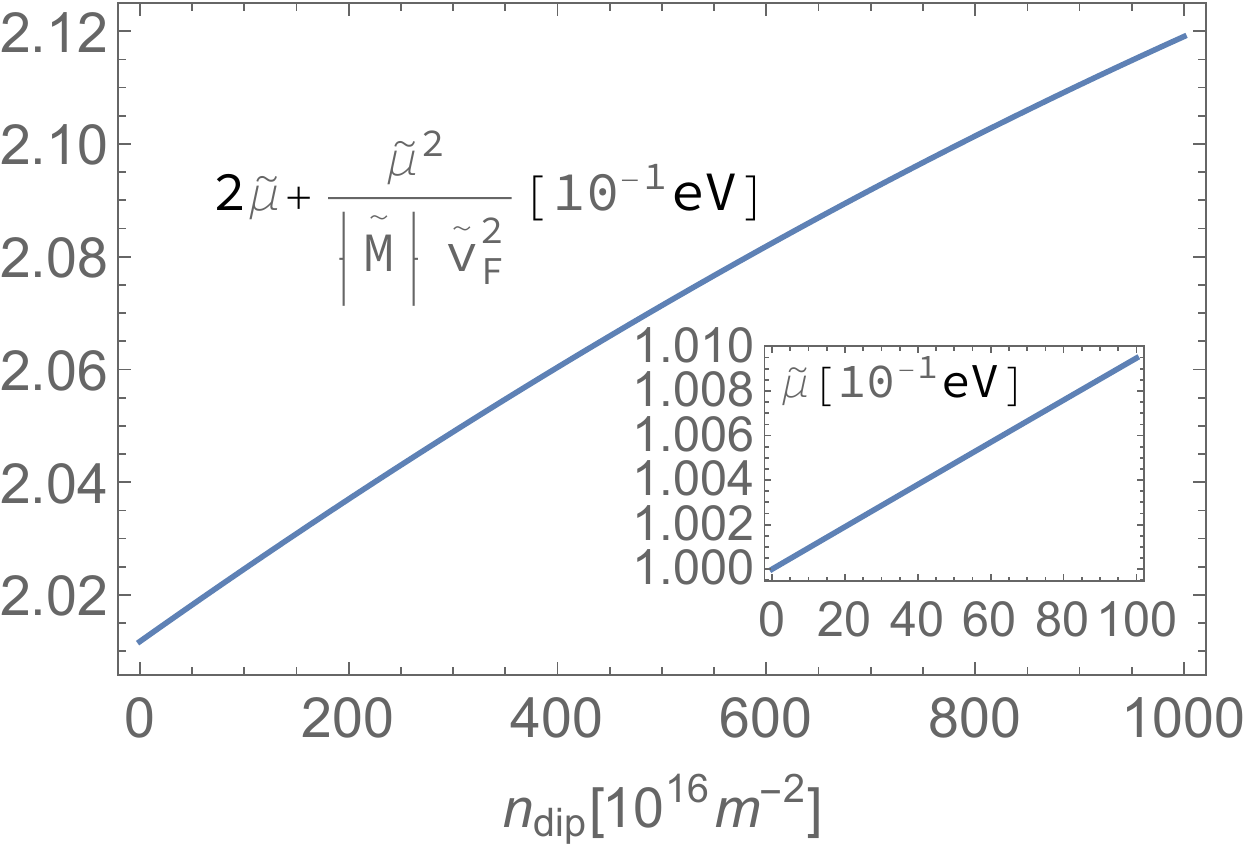}
    \caption{The locus of the center of the sigmoidal conductance curve (main figure),
    and (Inset) renormalized chemical potential,
    as a function of
    adsorbed molecular concentration $n_{dip}$. }
    \label{fig:chempot}
\end{figure}

Following the analysis in our previous work \cite{Falomir_2018,Falomir_2020}, we obtain 
that the tensor is symmetric and diagonal, $\sigma_{11} = \sigma_{22}$, $\sigma_{12} = \sigma_{21} = 0$. In particular,
for the real part of the optical conductivity that can be measured in electronic transport experiments, we obtain the
analytical expression \cite{Falomir_2020}
\begin{widetext}
\begin{eqnarray}
\Re\, \sigma_{11}(\omega,T) =
\frac{e^2}{8\hbar}\sgn(\omega)\left(
\tanh\left[\frac{1}{2 k_B T}\left(\frac{\hbar^2 \omega^2}{8 \tilde{M} \tilde{v}_F^2} + \frac{\hbar\omega}{2} - \tilde{\mu} \right)\right] - \tanh\left[\frac{1}{2 k_B T}\left(\frac{\hbar^2 \omega^2}{8 \tilde{M} \tilde{v}_F^2} - \frac{\hbar\omega}{2} - \tilde{\mu} \right) \right]
\right)
\label{eq:graphcond}
\end{eqnarray}
\end{widetext}
We notice that the zero temperature limit of the former expression reduces to the universal value $e^2/(4\hbar)$, as follows \cite{Falomir_2020}
\begin{eqnarray}
\Re\, \sigma_{11}(\omega,T\rightarrow 0) = \left\{
\begin{array}{cc}
\frac{e^2}{4\hbar}, &  \frac{\hbar|\omega|}{2|\tilde{M}|\tilde{v}_F^2} > 1 - \sqrt{1 - \frac{2\tilde{\mu}}{|\tilde{M}|\tilde{v}_F}},\nonumber\\
0, & {\text{otherwise}}
\end{array}
\right.
\end{eqnarray}
In Fig.~\ref{fig:cond}, we display the optical conductance,
normalized by the "universal" value $\sigma_0 = e^{2}/(4\hbar)$,
as a function of frequency for different molecular surface concentrations. For this example, we used $T = 30$~K and the
representative parameters of $\text{NH}_3$.
At finite temperature, the center of the sigmoidal
curve is located at $\hbar\omega = 2 |\tilde{M}|\tilde{v}_F^2\left(1 - \sqrt{1 - \frac{2\tilde{\mu}}{|\tilde{M}|\tilde{v}_F^2}}  \right)\sim 2\tilde{\mu} + \frac{\tilde{\mu}^2}{|\tilde{M}|\tilde{v}_F^2}$.
In Fig.~\ref{fig:chempot}, we display the locus of this point
as a function of the molecule surface concentration, for the same representative parameters of $\text{NH}_3$. The inset of this figure displays the corresponding values for the renormalized chemical potential. A change of about $10\%$
is predicted from the model, when the bare chemical potential is
chosen as $\mu = 0.1$~eV.

\section*{Conclusion}
Based on a quantum field theory description, we obtained a model for the optical conductivity of graphene
at different concentrations of adsorbed polar molecules. Our analytical results show that from electric transport measurements at finite
frequency, it is possible to read the shift
in the conductivity curve, and hence to infer the numerical
values of the corresponding renormalized parameters $\tilde{\mu}$, $\tilde{M}$ and $\tilde{v}_F$. By comparing these
values with the bare graphene parameters $M$ and $v_F$, as well as with the actual chemical potential $\mu$,
using the analytical Eq.~(\ref{eq:renorparam}) and Eq.~(\ref{eq:z})
the concentration of adsorbed molecules $n_{dip}$ can be estimated.

\appendix
\section{}\label{AppendixA}
In this Appendix, we present the basic considerations
involved in the disorder averaging that leads to the NCA approximation applied in this work. This analysis closely follows \cite{Rammer_86}.
We start by considering the potential experienced at site $\mathbf{r} \in \mathbb{R}^2$, due to the presence of a
collection of $N$ dipolar molecules adsorbed on the graphene
surface. The position of the center of mass of each molecule is at $(\mathbf{R}_j,a)$, with $a > 0$ the height with respect to the graphene surface (at $z = 0$) and $\mathbf{R}_j\in\mathbb{R}^2$ the position over the plane,
for $1 \le j \le N$. The corresponding potential is
\begin{eqnarray}
V\left(\mathbf{r} \right) = \sum_{j=1}^N V_{\boldsymbol{d}_{j}}\left( \mathbf{r} - \mathbf{R}_j \right),
\end{eqnarray}
where, as displayed in the main text, the potential for each dipole is 
\begin{eqnarray}
V_{\mathbf{d}_{j}}(\mathbf{r})
= \frac{\mathcal{P}}{4\pi\epsilon d}\left(\frac{1}{\sqrt{(\mathbf{r} - \frac{\mathbf{d}_j}{2})^2 + a^2}} -  \frac{1}{\sqrt{(\mathbf{r} + \frac{\mathbf{d}_j}{2})^2 + a^2}}\right)\nonumber\\
\end{eqnarray}
The 2D Fourier transform of this potential is given by the expression 
\begin{eqnarray}
\hat{V}_{\mathbf{d}_j}(\mathbf{q})
&=& \int_{\mathbb{R}^2} d^2 r e^{i\mathbf{q}\cdot\mathbf{r}}
V_{\mathbf{d}_j}(\mathbf{r})\nonumber\\
&=& \frac{i \mathcal{P} \sin\left(\frac{\mathbf{q}\cdot\mathbf{d}_j}{2} \right)}{2\pi\epsilon d}
\int_0^{\infty}dr \frac{r}{\sqrt{r^2 + a^2}}\int_0^{2\pi}d\phi e^{i q r \cos(\phi)}\nonumber\\
&=& \frac{i \mathcal{P} \sin\left(\frac{\mathbf{q}\cdot\mathbf{d}_j}{2} \right)}{\epsilon d}
\int_0^{\infty}dr \frac{r J_0(q r)}{\sqrt{r^2 + a^2}}\nonumber\\
&=& \frac{i \mathcal{P} \sin\left(\frac{\mathbf{q}\cdot\mathbf{d}_j}{2} \right)}{\epsilon q d}
e^{- q a}
\end{eqnarray}
In the strict dipole approximation, i.e. $d\rightarrow 0$ but $\mathcal{P}_{j}$ finite,
the Fourier transform reduces to the simpler expression
\begin{eqnarray}
\lim_{d\rightarrow 0} \hat{V}_{\mathbf{d}_j}(\mathbf{q}) \equiv
\hat{V}_{\boldsymbol{\mathcal{P}}_{j}}(\mathbf{q})
= \frac{i \mathbf{q}\cdot\boldsymbol{\mathcal{P}}_{j}}{2 \epsilon q }
e^{-q a}
\end{eqnarray}
With this result, the 2D Fourier transform of the potential is given by
\begin{eqnarray}
\hat{V}(\mathbf{q}) &=& \int_{\mathbb{R}^2} d^2 r e^{i \mathbf{q}\cdot\mathbf{r}}
V(\mathbf{r})\nonumber\\
&=& \sum_{j=1}^N \int_{\mathbb{R}^2}
d^2 r e^{i\mathbf{q}\cdot\mathbf{r}}V_{\boldsymbol{d}_{j}}(\mathbf{r}- \mathbf{R}_j)\nonumber\\
&=& \sum_{j=1}^N e^{i\mathbf{q}\cdot\mathbf{R}_j}\int_{\mathbb{R}^2}
d^2 r e^{i\mathbf{q}\cdot\mathbf{r}}V_{\boldsymbol{d}_{j}}(\mathbf{r})\nonumber\\
&=& \sum_{j=1}^{N}\hat{V}_{\boldsymbol{d}_{j}}(\mathbf{q})e^{i\mathbf{q}\cdot\mathbf{R}_j} \xrightarrow{d\rightarrow 0} \sum_{j=1}^{N}\hat{V}_{\boldsymbol{\mathcal{P}}_{j}}(\mathbf{q})e^{i\mathbf{q}\cdot\mathbf{R}_j}
\end{eqnarray}
We shall assume that the positions of the adsorbed molecules
over the graphene surface, i.e. the set $\{\mathbf{R}_j \}_{j=1}^{N}$, as well as the orientation of the dipole
moments $\{\boldsymbol{\mathcal{P}}_{j} \}_{j=1}^{N}$ are independent and identically distributed random variables. Therefore,
writing $\boldsymbol{\mathcal{P}}_{j} = \mathcal{P}\left(\cos(\varphi_j),\sin(\varphi_j) \right)$,
with $\varphi_j \in [0,2\pi]$ a uniformly distributed random variable
\begin{eqnarray}
\langle \boldsymbol{\mathcal{P}}_{j}\rangle
= \mathcal{P}  \left(\langle \cos(\varphi_j)\rangle,\langle \sin(\varphi_j)\rangle\right)  = 0
\end{eqnarray}
\begin{eqnarray}
\langle \hat{V}_{\boldsymbol{\mathcal{P}}_{j}}(\mathbf{q})\rangle
&=& \frac{i e^{-q a}}{2\epsilon q}\langle \mathbf{q}\cdot\boldsymbol{\mathcal{P}}_{j}\rangle\nonumber\\
&=& \frac{i e^{-q a}}{2\epsilon q}\mathbf{q}\cdot\langle \boldsymbol{\mathcal{P}}_{j}\rangle = 0.
\end{eqnarray}
Similarly, for $\mathbf{q}_1 = q_1\left(\cos(\phi_1),\sin(\phi_1)\right)$
and $\mathbf{q}_2 = q_2\left(\cos(\phi_2),\sin(\phi_2)\right)$
\begin{eqnarray}
&&\langle \hat{V}_{\boldsymbol{\mathcal{P}}_{j}}(\mathbf{q}_1) \hat{V}_{\boldsymbol{\mathcal{P}}_{j'}}(\mathbf{q}_2)\rangle
= \frac{i^2 e^{-(q_1 + q_2)a}}{4\epsilon^2 q_1 q_2}\nonumber\\
&&\times\langle \mathbf{q}_1\cdot\boldsymbol{\mathcal{P}}_{j}\mathbf{q}_2
\cdot\boldsymbol{\mathcal{P}}_{j'}\rangle\nonumber\\
&=& \frac{i^2 e^{-(q_1 + q_2)a} \mathcal{P}^2}{4\epsilon^2}
\left(
\cos\phi_1\cos\phi_2 \langle \cos\varphi_j\cos\varphi_{j'} \rangle\right.\nonumber\\
&&\left.
+ \sin\phi_1\sin\phi_2 \langle \sin\varphi_j\sin\varphi_{j'} \rangle + \cos\phi_1\sin\phi_2\right.\nonumber\\
&&\left.\times\langle
\cos\varphi_j\sin\varphi_{j'}\rangle
+ \sin\phi_1\cos\phi_2\langle
\cos\varphi_j\sin\varphi_{j'}\rangle
\right)\nonumber\\
&&=\frac{i^2 e^{-(q_1 + q_2)a}\mathcal{P}^2}{8 \epsilon^2}\delta_{j,j'}
\left(
\cos\phi_1\cos\phi_2 + \sin\phi_1\sin\phi_2
\right)\nonumber\\
&&=\delta_{j,j'}\frac{i^2 e^{-(q_1 + q_2)a}\mathcal{P}^2}{8\epsilon^2}\frac{\mathbf{q}_1\cdot\mathbf{q}_2}{q_1q_2}
\end{eqnarray}
Generalizing this procedure, it is straightforward to
show that the average of the product of an odd number
of potential terms is identically zero, since
\begin{eqnarray}
\langle \left(\cos(\varphi_j)\right)^{2n + 1} \rangle = 
\langle \left(\sin(\varphi_j)\right)^{2n + 1} \rangle = 0
\end{eqnarray}
Similarly, for the average over the molecular positions $\mathbf{R}_j$, we have
\begin{eqnarray}
\langle 
e^{i \left(\mathbf{q} - \mathbf{q}'\right)\cdot\mathbf{R}_j}
\rangle = \frac{1}{A}\int_{\mathbb{R}^2}d^2 R_j e^{i \left(\mathbf{q} - \mathbf{q}'\right)\cdot\mathbf{R}_j} = \frac{(2\pi)^2}{A}\delta\left(\mathbf{q} - \mathbf{q}' \right)
\nonumber\\
\end{eqnarray}
Let us now consider the Lippmann-Schwinger series, in momentum space, for the scattering process across a given realization of the ensemble of adsorbed molecules,
\begin{eqnarray}
\hat{G}(\mathbf{p},\mathbf{p'};\omega)
&=&(2\pi)^2\delta(\mathbf{p} - \mathbf{p}')\hat{\Delta}_{0}^{R}(\mathbf{p},\omega)\nonumber\\
&+& \hat{\Delta}_{0}^{R}(\mathbf{p},\omega)e \hat{V}\left(\mathbf{p}-\mathbf{p}' \right)\hat{\Delta}_{0}^{R}(\mathbf{p}',\omega)
\nonumber\\
&+& \int \frac{d^2 p''}{(2\pi)^2}\hat{\Delta}_{0}^{R}(\mathbf{p},\omega)
e \hat{V}\left(\mathbf{p}-\mathbf{p}'' \right)\hat{\Delta}_{0}^{R}(\mathbf{p}'',\omega)\nonumber\\
&&\times e \hat{V}\left(\mathbf{p}''-\mathbf{p}' \right)\hat{\Delta}_{0}^{R}(\mathbf{p}',\omega)+\ldots
\end{eqnarray}
We can directly verify that, after performing a statistical
average over the disordered distribution of positions and
dipole orientations, the translational invariance of the
Green's function is recovered. For instance, for the first-order term, we have
\begin{eqnarray}
\langle \hat{G}^{(1)}(\mathbf{p},\mathbf{p'};\omega)\rangle
=\hat{\Delta}_{0}^{R}(\mathbf{p},\omega)\langle \hat{V}\left(\mathbf{p}-\mathbf{p}' \right)\rangle\hat{\Delta}_{0}^{R}(\mathbf{p}',\omega)
\end{eqnarray}
Notice that the disorder average of the potential, under
the assumptions of statistical independence of the
parameters, is
\begin{eqnarray}
\langle \hat{V}\left(\mathbf{p}-\mathbf{p}' \right)\rangle
&=& \sum_{j=1}^{N}\langle e^{i(\mathbf{p} - \mathbf{p}')\cdot\mathbf{R}_j} \rangle \langle \hat{V}_{\boldsymbol{\mathcal{P}}_{j}}(\mathbf{p} - \mathbf{p}') \rangle\nonumber\\
&=&\frac{N}{A}(2\pi)^2 \delta(\mathbf{p} - \mathbf{p}')
\langle \hat{V}_{\boldsymbol{\mathcal{P}}_{j}}(0)\rangle\nonumber\\
&=& 0
\end{eqnarray}
Therefore, clearly this first correction vanishes after averaging over dipole orientations. For the second order contribution, we have
\begin{eqnarray}
\langle \hat{G}^{(2)}(\mathbf{p},\mathbf{p}';\omega)\rangle
&=& \int\frac{d^2 p''}{(2\pi)^2}\hat{\Delta}_{0}^{R}(\mathbf{p},\omega)\hat{\Delta}_0^{R}(\mathbf{p}'',\omega)\hat{\Delta}_0^{R}(\mathbf{p}',\omega)\nonumber\\
&&\times e^2\langle \hat{V}(\mathbf{p} - \mathbf{p}'')\hat{V}(\mathbf{p}'' - \mathbf{p}')\rangle
\end{eqnarray}
Here, the disorder average of the product of potential
factors is
\begin{widetext}
\begin{eqnarray}
\langle e^2 \hat{V}(\mathbf{p} - \mathbf{p}'')\hat{V}(\mathbf{p}'' - \mathbf{p}')\rangle &=& \sum_{j=1}^{N}\sum_{j'=1}^{N}
\langle e^{i(\mathbf{p} - \mathbf{p}'')\cdot\mathbf{R}_j}
e^{i(\mathbf{p}'' - \mathbf{p}')\cdot\mathbf{R}_{j'}}
\langle e^2 \hat{V}_{\boldsymbol{\mathcal{P}}_{j}}(\mathbf{p} - \mathbf{p}'')
\hat{V}_{\boldsymbol{\mathcal{P}}_{j'}}(\mathbf{p}'' - \mathbf{p}')\rangle\nonumber\\
&=& \sum_{j=1}^{N}\langle e^{i(\mathbf{p} - \mathbf{p}')\cdot\mathbf{R}_j} \rangle
\langle e^2 \hat{V}_{\boldsymbol{\mathcal{P}}_{j}}(\mathbf{p} - \mathbf{p}'')
\hat{V}_{\boldsymbol{\mathcal{P}}_{j}}(\mathbf{p}'' - \mathbf{p}')\rangle = \frac{N}{A}(2\pi)^2\delta(\mathbf{p} - \mathbf{p}')
\frac{e^2 \mathcal{P}^2}{8\epsilon^2}e^{-2a|\mathbf{p} - \mathbf{p}''|}\nonumber\\
&=& (2\pi)^2\delta(\mathbf{p} - \mathbf{p}')n_{dip}|\hat{V}_{avg}(\mathbf{p} - \mathbf{p}'')|^2
\end{eqnarray}
\end{widetext}
Therefore, from this result we obtain that the second order contribution is
\begin{eqnarray}
\langle\hat{G}^{(2)}(\mathbf{p},\mathbf{p'};\omega)\rangle
=(2\pi)^2\delta(\mathbf{p} - \mathbf{p}')\hat{\Delta}^{R}(\mathbf{p},\omega),
\end{eqnarray}
with
\begin{eqnarray}
&&\hat{\Delta}^{R}(\mathbf{p},\omega) = \hat{\Delta}^{R}_0(\mathbf{p},\omega)
+ \hat{\Delta}^{R}_0(\mathbf{p},\omega)\\
&&\times n_{dip}
\int \frac{d^{2}p''}{(2\pi)^2}|\hat{V}_{avg}(\mathbf{p} - \mathbf{p}'')|^2\hat{\Delta}^{R}_0(\mathbf{p}'',\omega)
\hat{\Delta}^{R}_0(\mathbf{p},\omega)+\ldots\nonumber,
\end{eqnarray}
where we defined the disordered-averaged potential energy
\begin{eqnarray}
|V_{avg}(\mathbf{p} )|^2 = \frac{e^2\mathcal{P}^2}{8\epsilon^2}e^{-2a|\mathbf{p}|}
\end{eqnarray}

\section{}\label{AppendixB}
In this Appendix, we present the calculation of the
self-energy coefficients $\mathcal{I}_{1}(i\omega_n)$
and $\mathcal{I}^{(2)}(i\omega_n)$, as defined in the
main text.

By defining the coordinates $\mathbf{p} = p\left(\cos\varphi,\sin\varphi \right)$, 
and $\mathbf{q} = q\left(\cos\left(\theta - \varphi \right),\sin\left(\theta - \varphi \right) \right)$, we have
\begin{eqnarray}
\mathbf{q}\cdot\boldsymbol{\sigma} = q \left(\begin{array}{cc}0 & e^{-i\left(\theta - \varphi \right)}\\ e^{i\left(\theta - \varphi \right)} & 0 \end{array}\right).
\end{eqnarray}
Therefore, the integrals involved in the calculation of the self-energy are of the forms
\begin{eqnarray}
\mathcal{I}_1(i\omega_n,\mathbf{p}) &=& \int\frac{d^2 q}{(2\pi)^2} \frac{e^{-2|\mathbf{p} - \mathbf{q}|a}\left(i\omega_n + \mu - \frac{q^2}{2 M}  \right)}{\left(i\omega_n + \mu - \frac{q^2}{2 M}  \right)^2
- v_F^2 q^2}\nonumber\\
\mathcal{I}_2(i\omega_n,\mathbf{p}) &=& \int\frac{d^2 q}{(2\pi)^2} \frac{e^{-2|\mathbf{p} - \mathbf{q}|a} q e^{\pm i \left(\theta - \varphi\right)}}{\left(i\omega_n + \mu - \frac{q^2}{2 M}  \right)^2
- v_F^2 q^2},
\end{eqnarray}
such that the self-energy is expressed as
\begin{eqnarray}
\hat{\Sigma}(i\omega_n,\mathbf{p}) = n_{dip}\frac{e^2\mathcal{P}^2}{8\epsilon^2}
\left(\mathbf{1}\,\mathcal{I}_1(i\omega_n,\mathbf{p}) + \frac{\boldsymbol{\sigma}\cdot\mathbf{p}}{p}\mathcal{I}_2(i\omega_n,\mathbf{p}) \right)\nonumber\\
\end{eqnarray}
where we use polar coordinates $d^2q = q dq d\theta$, $\theta \in [0,2\pi)$ and $q \in [0,\infty)$. To evaluate those integrals, we perform a series expansion in powers of the external momentum $\mathbf{p}$, retaining up to second order consistently with
the next-to-nearest neighbors approximation in the graphene Hamiltonian.
Therefore, for the first integral we have
\begin{equation}\label{L1}
    \mathcal{I}_1(i\omega_n,\mathbf{p}) 
      = \mathcal{I}_1^{(0)}(i\omega_n) + \mathbf{p}^2 \, \mathcal{I}_1^{(2)}(i\omega_n) + O\left(p^4\right),
\end{equation}
where
\begin{widetext}
\begin{equation}\label{L2}
    \begin{array}{c} \displaystyle
      \mathcal{I}_1^{(0)}(i\omega)=-\frac{\mu +i \omega }{8 \pi  {v_F}^2} \displaystyle
      \left\{
      e^{-\frac{2 a (\mu +i \omega )}{{v_F}}} \Gamma \left(0,-\frac{2 a (\mu +i
   \omega )}{{v_F}}\right)+e^{\frac{2 a (\mu +i \omega )}{{v_F}}} \Gamma
   \left(0,\frac{2 a (\mu +i \omega )}{{v_F}}\right)
    \right\}+ \frac{(\mu +i \omega )^2}{16 \pi  M {v_F}^5} \, \times
   \\ \\ \displaystyle
   \left\{
   \frac{{v_F} \left[{v_F}^2-4 a^2 (\mu +i \omega )^2\right]}{2 a^2 (\mu +i \omega )^2}+
   e^{\frac{2 a (\mu +i \omega )}{{v_F}}} [3 {v_F}+2 a (\mu +i \omega )]
   \Gamma \left(0,\frac{2 a (\mu +i \omega )}{{v_F}}\right)
   \right.
    \\ \\ \displaystyle
    \left.
    -e^{-\frac{2 a
   (\mu +i \omega )}{{v_F}}} [-3 {v_F}+2 a (\mu +i \omega )] \Gamma
   \left(0,-\frac{2 a (\mu +i \omega )}{{v_F}}\right)
       \right\}+O\left( \frac{1}{M^2}\right)
    \end{array}
\end{equation}
\end{widetext}
and
\begin{widetext}
\begin{equation}\label{L3}
    \begin{array}{c} \displaystyle
      \mathcal{I}_1^{(2)}(i\omega)=\frac{1}{8 \pi  {v_F}^2}\left\{\frac{{v_F}^2}{\mu +i \omega }
      -a e^{-\frac{2 a (\mu +i \omega )}{{v_F}}} [-{v_F}+2 a (\mu +i \omega )]
   \Gamma \left(0,-\frac{2 a (\mu +i \omega )}{{v_F}}\right)-
   a e^{\frac{2 a
   (\mu +i \omega )}{{v_F}}} [{v_F}+2 a (\mu +i \omega )]\right.\,\times \\ \\ \displaystyle
   \left.
   \Gamma
   \left(0,\frac{2 a (\mu +i \omega )}{{v_F}}\right) \right\}
   +\frac{a (\mu +i \omega )}{8 \pi  M {v_F}^5}
   \left\{-2 a {v_F} (\mu +i \omega ) +
   e^{\frac{2 a (\mu +i \omega )}{{v_F}}} \left[2 a^2 (\mu +i \omega )^2+4 a
   {v_F} (\mu +i \omega )+{v_F}^2\right] \Gamma \left(0,\frac{2 a (\mu
   +i \omega )}{{v_F}}\right)
   \right.
   \\ \\ \displaystyle
   \left.
   -e^{-\frac{2 a (\mu +i \omega )}{{v_F}}}
   \left[2 a^2 (\mu +i \omega )^2-4 a {v_F} (\mu +i \omega
   )+{v_F}^2\right] \Gamma \left(0,-\frac{2 a (\mu +i \omega
   )}{{v_F}}\right)
    \right\}+O\left( \frac{1}{M^2}\right)
   \end{array}
\end{equation}
\end{widetext}
For the second integral we get
\begin{widetext}
\begin{equation}\label{L4}
    \begin{array}{c}\displaystyle
    \mathcal{I}_2(i\omega_n,\mathbf{p}) \asymp \frac{ p}{8 \pi  {v_F}} \left\{
    \left[  \left(  \frac{1}{4M} \partial_a +\frac{v_F}{2} \right)\partial_a\partial_\mu -1
    \right] \int_0^\infty {d q} \, \frac{e^{-2 a q} }{\left(i \omega+\mu -\frac{q^2}{2 M}+ {v_F} q \right)}\right\}
       = p \, \mathcal{I}_2^{(1)}(i\omega_n) +O\left(p^3\right)
    \end{array}
\end{equation}
\end{widetext}
where
\begin{widetext}
\begin{equation}\label{L5}
    \begin{array}{c} \displaystyle 
    \mathcal{I}_2^{(1)}(i\omega)= -\frac{1}{4 \pi  {v_F}^2}+\frac{a (\mu +i \omega )}{4 \pi  {v_F}^3}
    \left\{
    e^{\frac{2 a (\mu +i \omega )}{{v_F}}} \Gamma \left(0,\frac{2 a (\mu +i
   \omega )}{{v_F}}\right)-e^{-\frac{2 a (\mu +i \omega )}{{v_F}}}
   \Gamma \left(0,-\frac{2 a (\mu +i \omega )}{{v_F}}\right)
    \right\}\, 
      \\ \\ \displaystyle -\frac{1}{8 \pi  M {v_F}^6}
       \left\{ -2 {v_F}^2 (\mu +i \omega )+a (\mu +i \omega )^2
      \left[e^{-\frac{2 a (\mu +i \omega )}{{v_F}}} [-3 {v_F}+2 a (\mu +i \omega )]
   \Gamma \left(0,-\frac{2 a (\mu +i \omega )}{{v_F}}\right)+
   \right. \right.
   \\ \\ \displaystyle
   \left.\left.
   e^{\frac{2 a (\mu +i \omega )}{{v_F}}} [3 {v_F}+2 a (\mu +i \omega )] \Gamma
   \left(0,\frac{2 a (\mu +i \omega )}{{v_F}}\right)
      \right]\right\}+O\left( \frac{1}{M^2}\right)
    \end{array}
\end{equation}
\end{widetext}
Here, $\Gamma(0,z)$ is the Incomplete Gamma function, while $E_1(z)$ and $E_i(z)$ are the Exponential Integral functions.

The quasiparticle parameters introduced in Eq.~(13) and Eq.~(14) in the main text are thus obtained from the expressions above, and correspond to
\begin{widetext}
\begin{eqnarray}
\mathcal{I}_1^{(0)}(0) &=& \frac{1}{32 \pi  M v_F^5}\left[\frac{v_F \left(v_F^2-4 a^2 \mu ^2\right)}{a^2}+2 \mu  e^{\frac{2 a \mu }{v_F}} \left(\mu  (2 a \mu +3 v_F)-2 M v_F^3\right) E_1 \left(\frac{2 a \mu }{v_F}\right)-2 \mu  e^{-\frac{2 a \mu }{v_F}} \left(\mu  (2 a \mu -3 v_F)\right.\right.\nonumber\\
&&\left.\left.+2 M v_F^3\right) \left(\text{Ei}\left(\frac{2 a \mu }{v_F}\right)+i \pi \right)\right]
\label{eq:I10}
\end{eqnarray}
\begin{eqnarray}
&&\frac{\partial\mathcal{I}_1^{(0)}}{\partial(i \omega)}(0) =
-\frac{1}{8 \pi  M v_F^6}\left[-e^{-\frac{2 a \mu }{v_F}} \left(\text{Ei}\left(\frac{2 a \mu }{v_F}\right)+i \pi \right) \left(-2 a^2 \mu ^3-2 a \mu  M v_F^3+6 a \mu ^2 v_F+M v_F^4-3 \mu  v_F^2\right)\right.\nonumber\\
&&\left.+e^{\frac{2 a \mu }{v_F}} \left(-2 a^2 \mu ^3+2 a \mu  M v_F^3-6 a \mu ^2 v_F+M v_F^4-3 \mu  v_F^2\right) E_1 \left(\frac{2 a \mu }{v_F}\right)-2 M v_F^4+5 \mu  v_F^2\right]
\label{eq:dI10}
\end{eqnarray}
\begin{eqnarray}
\mathcal{I}_1^{(2)}(0)&=&
\frac{1}{8 \pi  M v_F^5}\left[\frac{v_F \left(M v_F^4-2 a^2 \mu ^3\right)}{\mu }+a e^{\frac{2 a \mu }{v_F}} \left(\mu  \left(2 a^2 \mu ^2+4 a \mu  v_F+v_F^2\right)-M v_F^3 (2 a \mu +v_F)\right) E_1\left(\frac{2 a \mu }{v_F}\right)\right.\nonumber\\
&&\left.-a e^{-\frac{2 a \mu }{v_F}} \left(\mu  \left(2 a^2 \mu ^2-4 a \mu  v_F+v_F^2\right)-M v_F^3 (v_F-2 a \mu )\right) \left(\text{Ei}\left(\frac{2 a \mu }{v_F}\right)+i \pi \right)\right]
\label{eq:I12}
\end{eqnarray}
\begin{eqnarray}
\mathcal{I}_2^{(1)}(0)&=&
\frac{1}{8 \pi  M v_F^7}\left[-a \mu   \left(e^{\frac{2 a \mu }{v_F}} E_1\left(\frac{2 a \mu }{v_F}\right) \left(\mu  (2 a \mu +3 v_F)-2 M v_F^3\right)+e^{-\frac{2 a \mu }{v_F}}\left(\text{Ei}\left(\frac{2 a \mu }{v_F}\right)+i \pi \right)\right.\right.\nonumber\\
&&\times\left.\left.\left(\mu  (2 a \mu -3 v_F)+2 M v_F^3\right)\right)-2 v_F^2 \left(M v_F^2-\mu \right)\right]
\label{eq:I21}
\end{eqnarray}
\end{widetext}

\begin{acknowledgments} 
\emph{Acknowledgements}---
H.F. is partially supported by CONICET (Argentina), and acknowledge support from UNLP through Proy. Nro. 11/X909 (Argentina). M. L. acknowledges support from ANID/PIA/Basal (Chile) under grant FB082, Fondecyt Regular Grant No. 1190192 and Fondecyt Regular Grant No. 1170107.
E.M. acknowledges financial support from the Fondecyt Regular Grant No. 1190361, and from Project Grant ANID PIA Anillo ACT192023. E.M. and M.L also acknowledge 
financial support from Fondecyt Regular Grant No. 1200483.

\end{acknowledgments}

%
\end{document}